\documentclass[journal = jpclcd]{achemso}
\usepackage{graphicx}
\usepackage{amsmath}
\usepackage{xcolor}
\usepackage{amssymb}
\usepackage{enumerate}
\usepackage{multirow}
\usepackage{changepage}
\usepackage{appendix}
\usepackage{xr}
\makeatletter
\newcommand*{\addFileDependency}[1]{
  \typeout{(#1)}
  \@addtofilelist{#1}
  \IfFileExists{#1}{}{\typeout{No file #1.}}
}
\makeatother

\newcommand*{\myexternaldocument}[1]{
    \externaldocument{#1}
    \addFileDependency{#1.tex}
    \addFileDependency{#1.aux}
}

\myexternaldocument{SI}
\author{Guorong Weng}
\author{Vojt\v{e}ch Vl\v{c}ek}
\affiliation{Department of Chemistry and Biochemistry, University of California, Santa Barbara, 93106, U.S.A}
\email{vlcek@ucsb.edu}

\title{Quasiparticles and Band Transport in Organized Nanostructures of Donor-Acceptor Copolymers}


\begin{document}

\begin{tocentry}
    \centering
    \includegraphics[width=5cm]{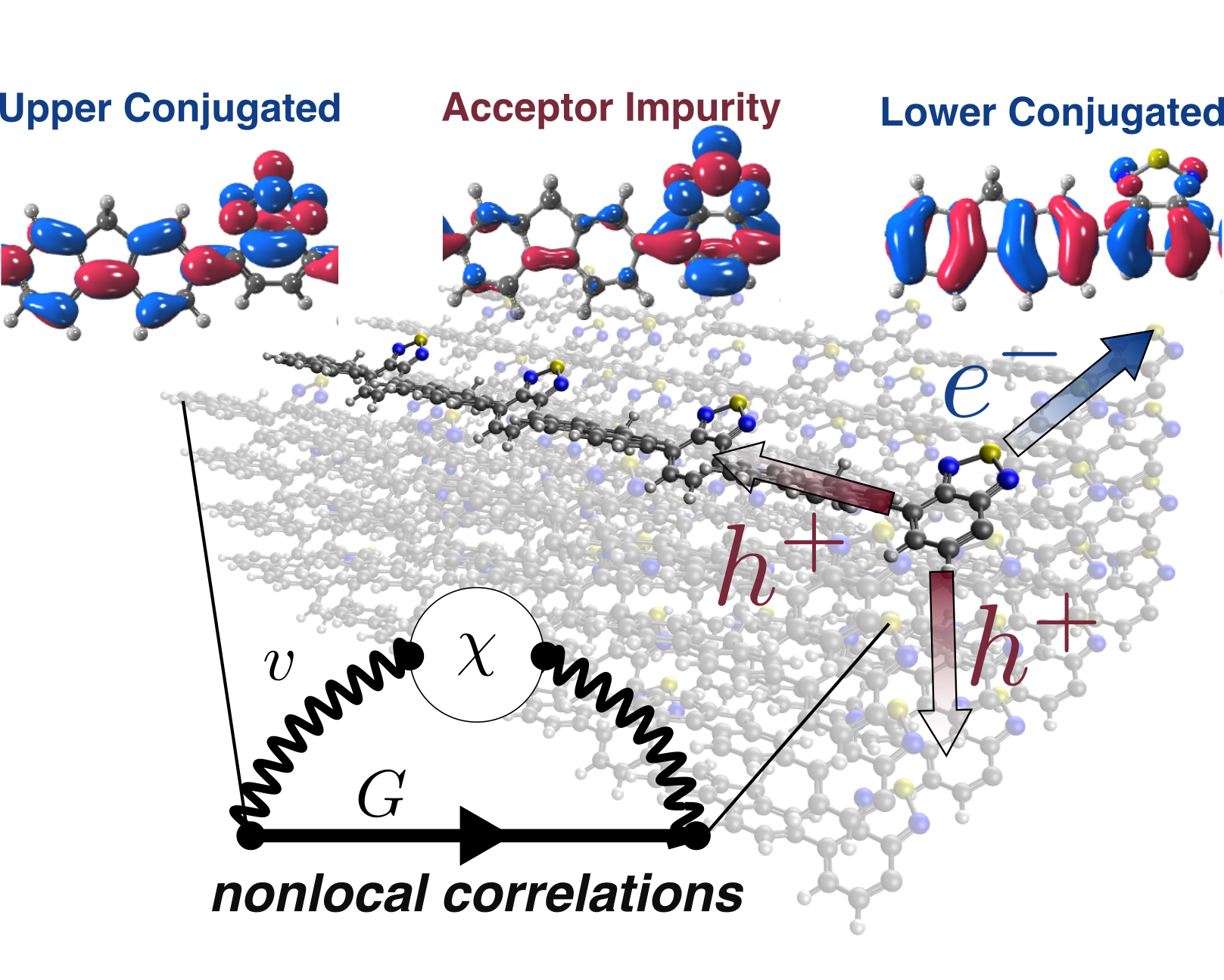}
    \label{fig:TOCentry}
\end{tocentry}

\date{}
\maketitle


\begin{abstract}

The performance of organic semiconductor devices is linked to highly-ordered nanostructures of self-assembled molecules and polymers. We employ many-body perturbation theory and study the excited states in bulk compolymers. We discover that acceptors in the polymer scaffold introduce a, hitherto unrecognized, conduction impurity band. The donor units are surrounded by conjugated bands which are only mildly perturbed by the presence of acceptors. Along the polymer axis, mutual interactions among copolymer strands hinder efficient band transport, which is, however, strongly enhanced across individual chains. We find that holes are most effectively transported in the $\pi-\pi$ stacking while electrons in the impurity band follow the edge-to-edge directions. The copolymers exhibit regions with inverted transport polarity, in which electrons and holes are efficiently transported in mutually orthogonal directions.

\end{abstract}


Donor-acceptor (D-A) semiconducting copolymers represent arguably the most variable class of semiconducting materials in organic electronics.\cite{Gunes2007,Heeger2010,Facchetti2011,Wang2012} The wide range of possible donors and acceptors provides an unmatched tunability of the system's electronic and optical properties.\cite{Yuen2011,Li2012,Zhang2013} Rational device-design is, however, hampered by the complicated relationship between electronic properties and the arrangement of the molecular chains in the condensed phase (e.g., in spin-coated thin-films).\cite{Schwartz2003,Noriega2013} Experiments showed that highly-ordered \emph{nanodomains}, i.e., highly organized nm-scale regions, are widely present in solution-processed thin films. The nanodomains are composed of nanowires,\cite{Oh2009} nanosheets,\cite{Barzegar2018} and crystallites.\cite{Sirringhaus1999,Venkateshvaran2014,Luo2014,Li2016} Face-on ($\pi$-$\pi$) or edge-on stacking is the dominant arrangement of conjugated molecules leading to high charge mobilities and excellent device performance.\cite{Oh2009,Barzegar2018,Sirringhaus1999,Venkateshvaran2014,Luo2014,Li2016} The nanodomains exhibit quasiparticle bands observed by angle-resolved photoemission \cite{Hsu2015}. Hence, the high hole mobilities are explained by band-like transport\cite{Sakanoue2010,Yamashita2014} in the $\pi$-$\pi$ direction\cite{Sirringhaus1999,Luo2014,Kim2014}. However, a detailed microscopic understanding of how the structure and composition of the copolymers impact the electronic excitations is currently missing.

Answering these questions requires a theoretical investigation of the copolymers' electronic structure in the condensed phase. In principle, such simulations need to capture the non-local inter-molecular interactions \cite{Sutton2016} of electrons delocalized along the $\pi$-conjugated backbone.\cite{Hsu2015} The individual polymer chains are highly polarizable and held together by van der Waals (vdW) forces. Even in the limit of ideally crystalline systems, quantitative theoretical predictions of electronic excitations are prohibitive, and they have been limited to crystals of small molecules\cite{Norton2008,Nayak2009,Difley2010,Ryno2013,Refaely-Abramson2013,Poelking2015,Kang2016,Li2016_Blase,Sun2016,Li2018,Bhandari2018}. For polymers, the computational efforts have considered only isolated\cite{Halls1999,Cornil2003} oligomers or 1D periodic systems\cite{Bredas1984,Bredas1985,cheng2017,He2018,Bredas2018} treated by mean-field approaches, which are less expensive but do not take into account the non-local electronic correlations (governed by polarization effects). \cite{Woods2016} Further, the geometries  of the polymer strands are typically forced to be planar, i.e., they disregard actual arrangements in the highly organized domains.\cite{Cornil2003,He2018} Finally, the mean-field methods do not, in principle, provide access to quasiparticle (injected electron and hole) energies and tend to underestimate excitation energies grossly.\cite{martin_reining_ceperley_2016}

In this work, we overcome these limitations and apply state-of-the-art theoretical approaches to explain the key features of the electronic structure of D-A copolymers.
Our calculations employ many-body perturbation theory\cite{martin_reining_ceperley_2016} within the stochastic $GW$ approach.\cite{Neuhauser2014,Vlcek2017,Vlcek2018,Vlcek2019} The electron-electron interactions are computed for each excitation (i.e., no mean-field approximation is applied). In the $GW$ approximation, the interaction term takes into account a selected class of Feynman diagrams describing the electrodynamic screening, i.e., the induced charge density fluctuations. Electrons thus interact via a screened Coulomb interaction, which is non-local and time-dependent. In practice, the $GW$ method yields quasiparticle (QP) excitation energies in excellent agreement with available experimental data.\cite{Blase2011,martin_reining_ceperley_2016,Vlcek2017}

\begin{figure}
    \centering
    \includegraphics[width=\linewidth]{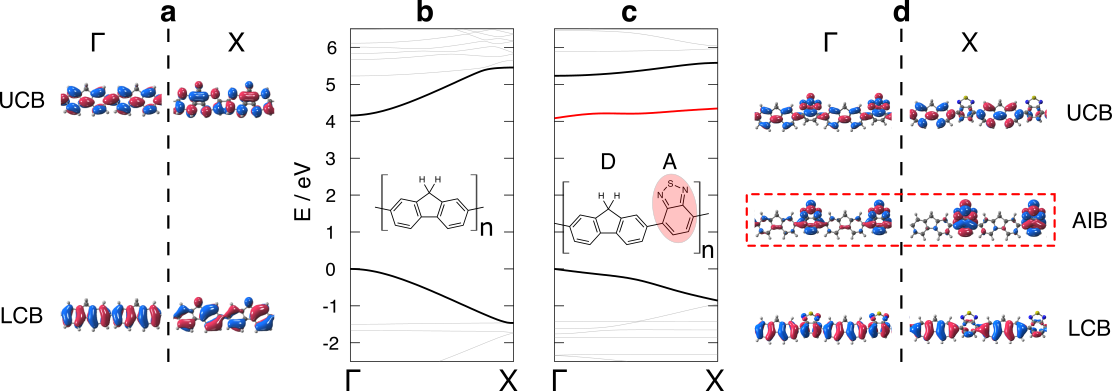}
    \caption{Quasiparticle band structures and orbitals of selected states of a fluorene (a, b) and FBT (c, d) strands. The monomer units are shown in the inset of panels b and c. The electronic states that are delocalized over the entire polymer backbone are denoted lower and upper conjugated bands (LCB and UCB), for the highest valence and lowest conduction band. The band-edge states in fluorene (a) are formed by LCB and UCB illustrated for the Brillouin zone center, $\Gamma$,  and its boundary, X. The corresponding bands are highlighted in panel b. FBT is D-A copolymer, with the individual subunits labeled in the inset of panel c. Due to the presence of A, the bandstructure contains an acceptor impurity band (AIB) highlighted in red (c). Panel d depicts LCB, AIB, and UCB for two points in the Brillouin zone; the AIB is strongly localized on the acceptor subunit. Red and blue colors distinguish the wave function phase.} 
    \label{Fig1Dbstr}
\end{figure}

The electronic structure and QP energies of the condensed phase is determined by the properties of the constituting moieties as well as by mutual interactions among individual copolymer strands.  While these contributions are nontrivial, relations among a few key parameters govern the system's overall behavior. To illustrate this, we consider a prototypical example:  ``FBT'' and related D-A copolymers\cite{Mai2013,Mai2015,Cui2018} (see Supporting Information (SI) for the geometry optimization). Here, the fluorene moiety (F) acts as a ``donor'' (D), and benzothiadiazole (BT) acts as an ``acceptor'' (A). The isolated molecules are illustrated in the inset of \textbf{Figure~\ref{Fig1Dbstr}} and the SI. The D units are the source of delocalized electronic states. In contrast, acceptors are typically chosen so that they have a higher electron affinity than donors\cite{Zhou2012,Duan2012}, acting as strong potential wells for electrons (see \textbf{Figure~\ref{Fig_IPEA}}).  Hence, the A unit is a source of localized electrons whose wave functions have a limited spatial extent. 

In a single copolymer strand (i.e.,  1D periodic system with repeated D and A subunits),  the quantum confinement is reduced in the direction of the polymer axis. Consequently, the fundamental gap of the infinite chain decreases with the polymerization length; for an infinite system, it is $4.08 \pm0.04$~eV, which is 1.48$\pm0.05$~eV less than for an isolated monomer (\textbf{Figure~\ref{Fig_EnergyDiagram}}).  In a condensed phase (either 2D slab or 3D bulk), the presence of neighboring strands eliminates the quantum confinement in the directions orthogonal to the polymer axis. Hence, the fundamental band gap  further decreases (Figure ~\ref{Fig_EnergyDiagram}b).

For quantitative predictions of the band gaps in the condensed phase, many-body methods turn out to be indispensable as dynamical electron-electron interactions are responsible for the non-local (inter-chain) interactions. Indeed, the ionization potential for the 2D slab computed with the s$GW$ method is 5.48$\pm0.02$~eV (Figure~\ref{Fig_EnergyDiagram}a), in excellent agreement with thin-film experiments that provide an estimate of 5.4-5.5 eV.\cite{Mai2015} The fundamental band gaps of the surface and the bulk are 3.33 and 2.23 eV (Figure~\ref{Fig_EnergyDiagram}b), and the latter is in good agreement with the experimental value of 2.32-2.44 eV.\cite{Mai2015}

The periodic copolymer arrangement supports the formation of band structures (observed experimentally, as discussed above). To characterize the principal features of the electronic states, we start with the 1D system shown in Figure~\ref{Fig1Dbstr}c. The crystal momentum is imprinted on the individual wavefunctions (Figure~\ref{Fig1Dbstr}d), which, however, retain much of their molecular character  (\textbf{Figure~\ref{Fig_Hybri}}). It is thus possible to separate the contributions of D and A  to the highest valence and lowest conduction bands responsible for the charge transport. 

The donor behavior dominates the highest valence state; it has conjugated character and delocalized $\pi$ orbitals (see more details in \textbf{Figure~\ref{Fig_1Dorbitals}}).  The top valence band is broad (its bandwidth is $0.86\pm0.04$~eV) with a parabolic dispersion near the extrema that occur at the critical points of the Brillouin zone. The near-band-edge character, together with the large bandwidth, translates to a low effective mass of $\sim0.22 m^*_e$. Such a low value is consistent with experimental results for similar (semi)conducting copolymers.\cite{Hsu2015}  We denote the highest conduction band and the lower conjugated band (LCB). The complementary ``upper'' conjugated band (UCB) is formed from $\pi^*$ orbitals, and it has much higher energy  (Figure~\ref{Fig1Dbstr}d). Both LCB and UCB are qualitatively analogous to the  band edge states in a \emph{pure} fluorene chain (Figure~\ref{Fig1Dbstr}a), i.e., the conjugated bands are only mildly perturbed by the presence of acceptor subunits. The correspondence between the electronic structures of D-A and pure donor polymers has not been noticed up to now.

In contrast, the lowest conduction band of the copolymer comprises states localized only on the acceptors (Figure~\ref{Fig_1Dorbitals}).  The acceptor band has significantly reduced width (Figure~\ref{Fig1Dbstr}c), and it appears between LCB and UCB.

In calculations with distinct A molecules, we found that the exact energy separation between the conjugated and localized states depends only a little on the choice of acceptors (see \textbf{Figure~\ref{Fig_AIB}} for details). In FBT, the separation of the conduction states is 1.11 eV; the oxygen- and selenium- substituted copolymers show slightly larger separations (1.38 eV and 1.16 eV, respectively -- see Figure~\ref{Fig_AIB}). In all cases studied, the localized state is characteristically inserted between the two conjugated bands. Based on the conceptual analogy to charge-trapping  ``in-gap'' states, we denote the lowest conduction states as the acceptor impurity band (AIB).  The formation of the localized and flat impurity band has not been described previously. One of the key findings of this communication is the recognition and distinction between the conjugated and impurity bands.

LCB, AIB, and UCB are present in the same order in 1D and in the condensed phases. While the van der Waals forces only weakly bond the individual copolymer strands, the inter-chain interactions change the band structures significantly. Besides the shift of the QP gaps (discussed above), the charge transport is critically influenced by the changes in the bandwidth. The dispersion of LCB and AIB  determines the charge transport polarity. Further,  the bandwidth is directly related to the charge carrier effective mass. To investigate the physical origin of the of the band structure changes, we will separate two main contributions: (i) the one-body electronic interactions\cite{onebodyterm} including the (classical) density-density Coulomb repulsion (\textbf{Table~\ref{tab:energydecomposition}}), and (ii) the electron-electron interactions,  which represent highly non-local and dynamical (time-dependent) quantum effects. 

The first contribution mostly depends on the local\cite{localproperties} properties of the copolymer. The electronic structure (and charge transport) strongly depend on the bond arrangement between the donor and acceptor subunits.\cite{Bredas1985}  The existence of a single bond between adjacent donors and acceptors implies large rotational freedom.  In practice, the mutual orientation of the A and D units depends on the environment. The rotational angle varies between 43$^\circ$ and 56$^\circ$ in the relaxed structures with 1D, 2D, or 3D topology (\textbf{Figure~\ref{Fig_torsion}}). Other structural variations are insignificant as the rest of the copolymer backbone is rigid, and we disregard them in the analysis.

\begin{figure}
    \centering
    \includegraphics[width=0.5\linewidth]{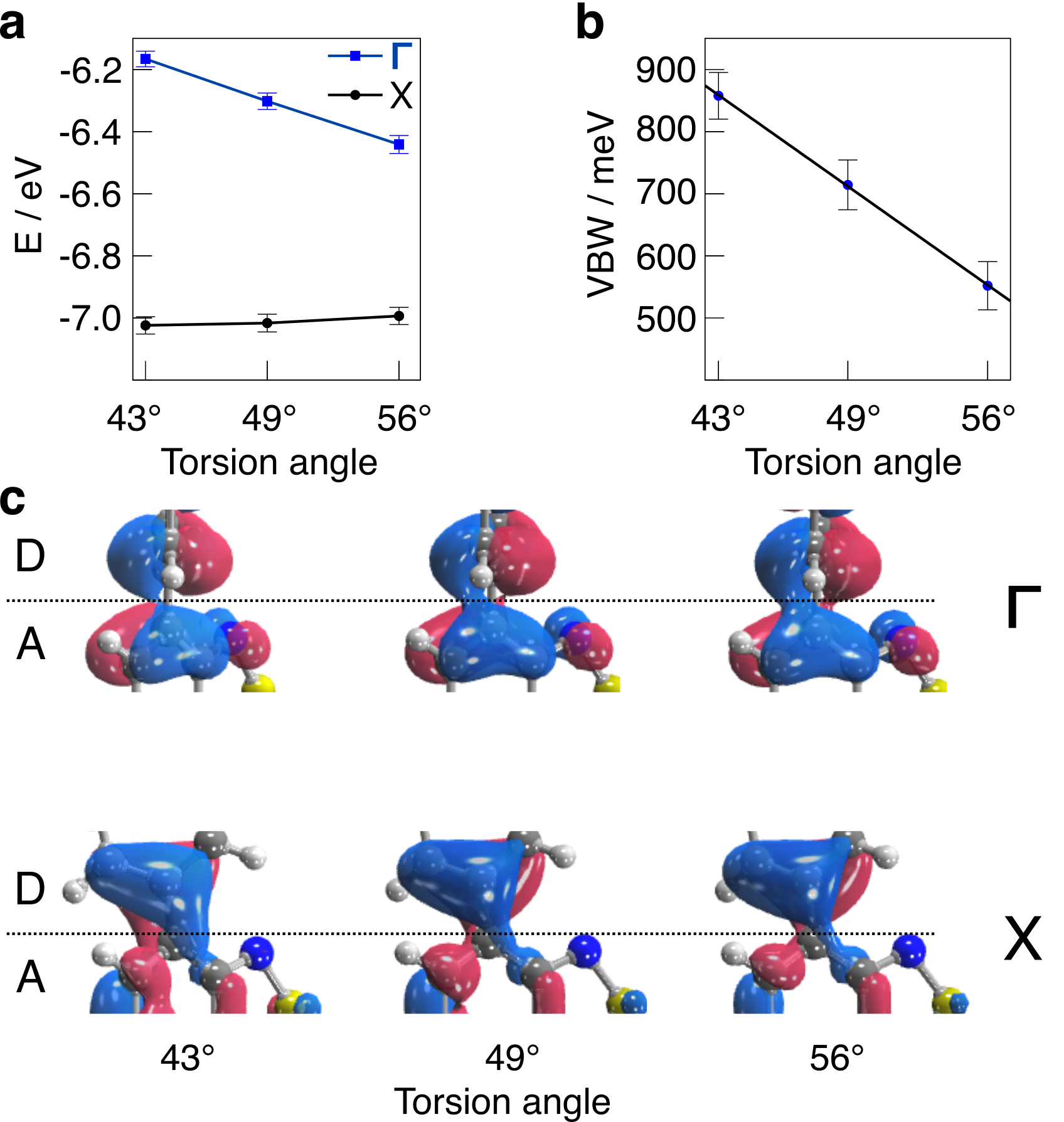}
    \caption{The effect of the characteristic torsion angles between D and A subunits found in various condensed phases of FBT: (a) The QP energies of LCB at the Brillouin zone center, $\Gamma$, and its boundary, X as a function of the torsion angle. The QP energy of LCB is more sensitive to torsion at $\Gamma$   than that at X. (b) The QP valence bandwidth linearly decreases with the torsion angle. (c) The LCB wavefunction at the donor (D) and the acceptor (A) joint for $\Gamma$ and $X$ points of the Brillouin zone. Red and blue colors distinguish the wave function phase. At $\Gamma$, the torsion gradually destroys a nodal plane between D and A, leading to kinetic stabilization of the QP energy. Conversely, the ``bridging character'' of LCB at the X point is little affected by the increased torsion. The error bars in panels a and b represent the statistical error of the stochastic many-body calculation.}
    \label{Figtorsion}
\end{figure}

As noted above, AIB is composed of  localized states  centered on the acceptor subunits. The corresponding wave function near the conduction edge does extent to the D-A joint appreciably (Figure~\ref{Fig1Dbstr}d). Hence, AIB is practically insensitive to the torsion angle.  In contrast, rotation away from the ideally planar geometry leads to the narrowing of conjugated states  (\textbf{Figure~\ref{Figtorsion}}b). Since the torsion angle is larger in the condensed phase than in a free-standing polymer, the hole effective mass in LCB is thus increased in bulk compared to a prediction from the 1D model.

The sensitivity of LCB is directly related to the character of the wave function near the D-A bond. Going from the low energy part of the LCB (near the X point of the Brillouin zone) to the band edge, the wavefunction develops a nodal plane across the D-A joint (Figure~\ref{Figtorsion}c). The presence of the nodes is associated with increased QP kinetic energy. A close inspection of various torsion angles reveals that the nodes across the D-A bond are suppressed when going from 1D to 3D conformation. The band edge is kinetically stabilized (Figure~\ref{Figtorsion}a), while the bottom LCB is insensitive to the rotation. As a result, the single-electron interactions promote bandwidth reduction in the condensed phase.

While the local properties are clearly responsible for the electronic structure modification, the non-local many-body effects are equally important and influence the excited states. These electron-electron interactions are decomposed into two principal contributions: (i) non-local exchange (due to the fermionic nature of the charge carriers), and (ii) time-dependent correlations among electrons and holes (which include vdW interactions responsible for the cohesive energy of the bulk).
The significance of the many-body treatment is illustrated by  the fact that LCB and AIB widths increase by $\sim 25\%$ and $\sim 46\%$  if the non-local and dynamic description is used instead of the common mean-field approach (e.g., in local and static density functional theory -- see \textbf{Table~\ref{tab:energydecomposition}}). 

We first inspect the behavior of the conjugated states. While the exchange interaction typically drives electron localization,\cite{Sanchez2008} it surprisingly enhances the dispersion of the delocalized bands along the polymer axis. The energies of states near the valence band maximum are stabilized much less than at the Brillouin zone boundary,  i.e., the X-point (\textbf{Figure~\ref{Fig_1DExP}}a). In the latter case, there is an increased spatial overlap with a large number of occupied orbitals, and energy decrease is observed for states near the X-point. The exchange-driven band widening is a signature of the conjugated bands, and it is not observed otherwise. To document this, we provide complementary calculations for additional polymer strands (polyacetylene and polyethylene, with and without conjugated bonds) in the SI (\textbf{Table~\ref{tab:exchange-driven_band_broadening}}). 

In general, this effect is dramatic for copolymer systems. In the absence of electronic correlation (which reduces the exchange through dynamical screening), LCB would widen by an additional 40\%. This increase can be paralleled with a (spurious) infinite-range response to hole localization observed for bare exchange interactions.\cite{Vlcek2016}

The screening contribution thus changes the picture qualitatively. It is governed by the reducible polarizability which is directly related to charge density fluctuations.\cite{martin_reining_ceperley_2016}
These correlation effects are dominated by optical (plasmon) excitation that shifts to lower energy as the crystal momentum increases (\textbf{Figure~\ref{Fig_CorrQPE}}). The states away from the band edge (i.e., closer to the Brillouin zone boundary) have energies approaching the resonant frequency of the collective charge density oscillations.  For the corresponding quasiparticle excitations, the exchange interaction is strongly attenuated and becomes short range; the QP energies shift up, and the LCB consequently narrows (Figure~\ref{Fig_1DExP}b).

\begin{figure}
    \centering
    \includegraphics[width=.99\textwidth]{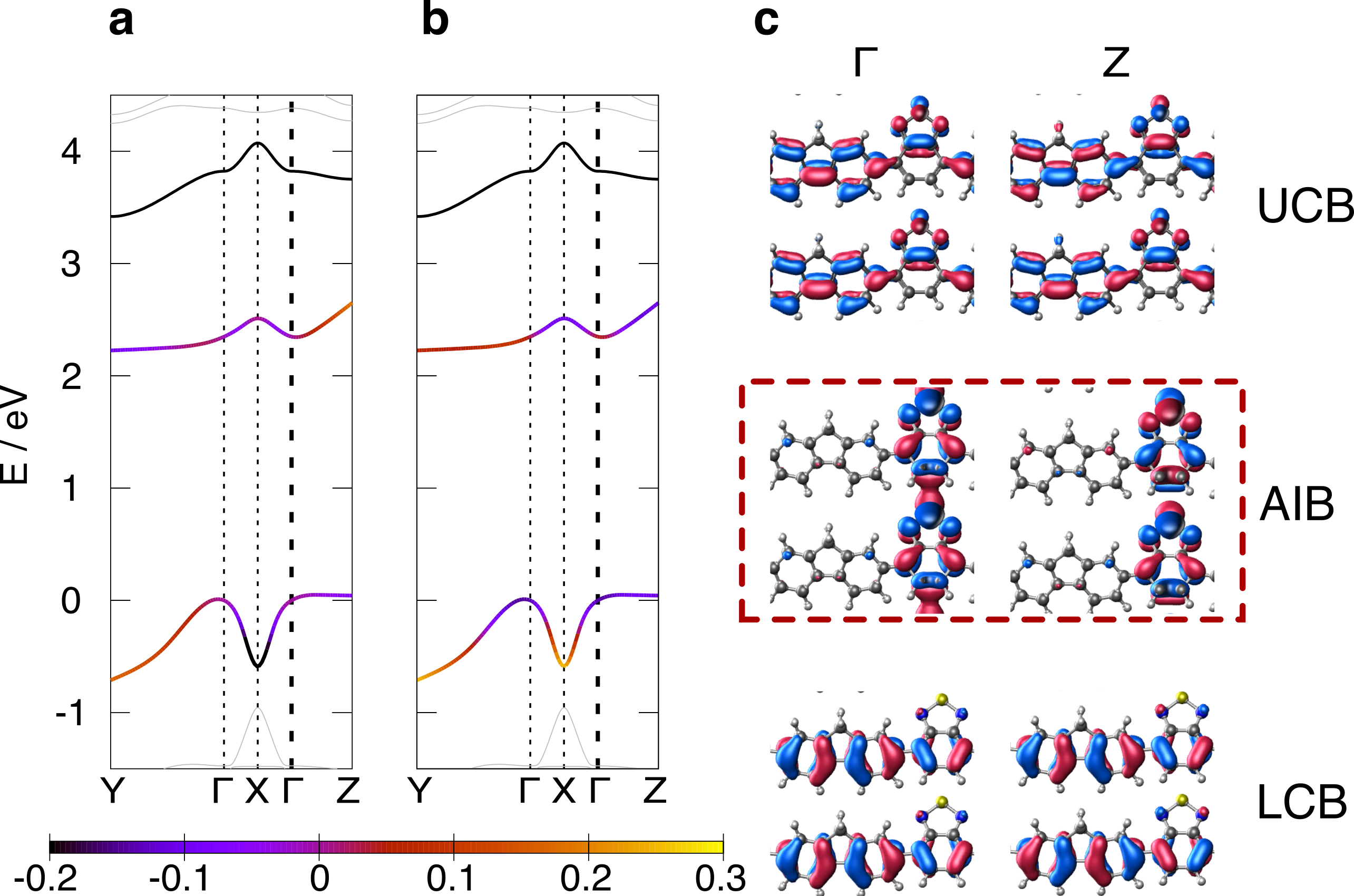}
    \caption{Quasiparticle band structure of FBT in a 3D crystalline domain with relative contributions of electronic (a) exchange and (b) correlation energies for LCB and AIB.  The contributions are given by the color code and are plotted relative to the band average. The $\Gamma \to$~Y branch corresponds to the band in the $\pi$-$\pi$ stacking direction; the $\Gamma \to$~X and $\Gamma \to$~Z branches correspond to the intra-chain and edge-on stacking directions. The inverted polarity regime is in the $\Gamma\to$~Z, where the band dispersion of AIB is much higher than for the conjugated bands. (c) Local wave function character is of the selected states at two distinct points in the Brillouin zone along the $\Gamma\to$~Z direction. The two molecules are depicted in the edge-on stacking. Both LCB and UCB remain localized on individual strands, but AIB bridges the polymer chains. Red and blue colors distinguish the wave function phase. }
    \label{Fig3Dbstr}
\end{figure}

In the condensed systems, the LCB and UCB  remain delocalized only along the polymer, not across the individual strands (illustrated in \textbf{Figure~\ref{Fig3Dbstr}}c). As a result, the conjugated bands can further flatten. Along the edge-to-edge direction ($\Gamma \to$Z Figure~\ref{Fig3Dbstr}a), both LCB and UCB are extremely narrow and effectively ``molecular'' in nature. Neither non-local exchange nor correlation effects contribute significantly to the quasiparticle energies in this case. In practice, any band-transport of holes in LCB is significantly hampered  along the edge-to-edge stacking direction. 

However, the localization does not imply that the conjugated bands behave as those in an isolated strand.  Here, the band dispersion is reduced by as much as 60\% along the polymer axis compared to a free-standing copolymer. The flattening is most prominent in the 2D case (\textbf{Table~\ref{tab:bandwidths1}}). Non-local interchain correlations govern the decrease of the LCB width; they are almost twice as big as the effect of torsion between the D and A subunits. In slabs, the strong polarization effects lead to the formation of local maxima in LCB and dispersion narrowing near the $\Gamma$ point (\textbf{Figure~\ref{Fig_2DExP}}b). This indicates that in near-surface regions, the valence band-edge may not be characterized by a single crystal momentum vector, and the fundamental band-gap is likely indirect.

In contrast, cooperative interchain interactions appear along the $\pi-\pi$ stacking  ($\Gamma \to$Y in Figure~\ref{Fig3Dbstr}). As a result, the LCB dispersion is the largest along this face-on direction  (up to 710~meV). Significant bandwidth suggests a high propensity for efficient band-like transport of holes within LCB. The reason for the increased bandwidth (despite the strong on-chain localization) is twofold: first, the packing of chains in bulk is tighter; second, the high efficient screening allows greater delocalization of the $\pi$ (and $\pi^*$) orbitals above and below the conjugated framework. Both effects lead to improved interchain ``communication'', which leads to an enlarged bandwidth. In the 3D condensed phase, the LCB width is largest along the $\pi-\pi$ stacking compared to any other direction (Figures~\ref{Fig3Dbstr}a and \ref{Fig3Dbstr}b) and indicates an efficient band-transport of holes.

The lowest conduction band is very different. The impurity states are strongly localized along the copolymer axis. As a result, local and non-local interactions are insensitive to crystal momentum, and the band is narrow. Further, there are no increased interactions along the $\pi-\pi$ stacking, and the AIB electronic states thus appear to be molecule-like. Hence, the acceptor band is  flat along $\Gamma\to$~Y as well. There is thus a low likelihood of electron band-transport in the face-on or polymer axis directions. 

AIB, however, unexpectedly exhibits cooperative effects along the edge-to-edge stacking (Figure~\ref{Fig3Dbstr}c). For states near the band minimum (i.e., near $\Gamma$), the impurity wavefunction delocalizes across individual copolymer chains. In contrast, a nodal plane appears between every adjacent polymer for higher crystal momenta due to the increase of the kinetic energy towards the Brillouin zone boundary (i.e., near Z).  The associated QP energy variation leads to a relatively wide\cite{dispersion} dispersion of $\sim$300 meV in the $\Gamma\to$Z direction. Besides the kinetic contribution, the band widening is also driven by a large variation of the exchange energy. Along $\Gamma\to$Z,  the AIB thus behaves like the conjugated states in the polymer axis. These properties indicate that AIB can sustain electron transport along the edge-to-edge stacking direction.

In summary, we investigated a prototypical example of D-A copolymers, FBT, and explained its electronic structure and the propensity to band transport in the condensed phase. Our many-body calculations are in excellent agreement with available experimental data and, for the first time, they provide insight into the quasiparticle (added hole and electron) states of bulk copolymers. The results show that acceptors, which typically act as strong potential wells for electrons, form a previously unrecognized ``impurity'' band. In contrast, the donors groups are responsible for delocalized lower (valence) and upper (conduction) conjugated bands. The delocalized states and they surround the acceptor band, but they only mildly affect each other.

The intra-chain transport is negatively impacted by the condensed phase stacking, which affects the rotation between the donor and acceptors. On the other hand, electronic states delocalize across the copolymer strands and form wide bands that likely support efficient transport. Electronic correlations (responsible for the cohesive van der Waals forces) universally suppresses band dispersion, but non-local exchange interactions drive it in selected directions. 

The large width of valence bands along the $\pi-\pi$ stacking indicates that hole transport is possible in the face-on direction. Surprisingly, we observe a strong propensity for electron transport along the edge-on stacking within the acceptor impurity band. Hence, D-A copolymers exhibit an orthogonal ambipolar transport network, which has been so far reported only in heterogeneous mixtures of p-type polymer and a small n-type molecule.\cite{Huang2016,Huang2018} Our results suggest that orthogonal transport of electrons and holes can be achieved in pure D-A copolymers merely through molecular packing.

\begin{suppinfo}
The Supporting Information provides additional texts, figures and tables listed below.

Texts: computational methods and details.

Figures: 1D, 2D, and 3D supercells in computations, QP energy diagrams of molecular and periodic systems, hybridization of FBT frontier orbitals, selected orbitals of UCB, AIB, and LCB in the 1D system, Comparison of band structures of D-A copolymers with different acceptors, QP band structures with exchange and correlation energy as functions of the momentum, graphical solutions to the QP and correlation energies, represnetaion of the D-A torsion angle.

Tables: parameters in DFT and MBPT calculations, measurements of geometrical constants for different polymers, decomposition of the contribution to the valence bandwidth, bandwidths of the LCB and AIB, exchange contribution to the valence bandwidth, measurements of torsion angle for FBT strands, convergence of the IP, EA, and gap to the supercell's size.

\end{suppinfo}

\section{Acknowledgement}
The authors want to acknowledge Prof.~Thuc-Quyen Nguyen and Prof.~Guillermo Bazan for fruitful discussions. This work was supported by the NSF CAREER award through Grant No. DMR-1945098. The calculations were performed as part of the XSEDE computational Project No. TG-CHE180051. Use was made of computational facilities purchased with funds from the National Science Foundation (CNS-1725797) and administered by the Center for Scientific Computing (CSC). The CSC is supported by the California NanoSystems Institute and the Materials Research Science and Engineering Center (MRSEC; NSF DMR 1720256) at UC Santa Barbara.

\bibliography{MB_vdW_OC}

\providecommand{\latin}[1]{#1}
\makeatletter
\providecommand{\doi}
  {\begingroup\let\do\@makeother\dospecials
  \catcode`\{=1 \catcode`\}=2\doi@aux}
\providecommand{\doi@aux}[1]{\endgroup\texttt{#1}}
\makeatother
\providecommand*\mcitethebibliography{\thebibliography}
\csname @ifundefined\endcsname{endmcitethebibliography}
  {\let\endmcitethebibliography\endthebibliography}{}
\begin{mcitethebibliography}{58}
\providecommand*\natexlab[1]{#1}
\providecommand*\mciteSetBstSublistMode[1]{}
\providecommand*\mciteSetBstMaxWidthForm[2]{}
\providecommand*\mciteBstWouldAddEndPuncttrue
  {\def\EndOfBibitem{\unskip.}}
\providecommand*\mciteBstWouldAddEndPunctfalse
  {\let\EndOfBibitem\relax}
\providecommand*\mciteSetBstMidEndSepPunct[3]{}
\providecommand*\mciteSetBstSublistLabelBeginEnd[3]{}
\providecommand*\EndOfBibitem{}
\mciteSetBstSublistMode{f}
\mciteSetBstMaxWidthForm{subitem}{(\alph{mcitesubitemcount})}
\mciteSetBstSublistLabelBeginEnd
  {\mcitemaxwidthsubitemform\space}
  {\relax}
  {\relax}

\bibitem[G{\"u}nes \latin{et~al.}(2007)G{\"u}nes, Neugebauer, and
  Sariciftci]{Gunes2007}
G{\"u}nes,~S.; Neugebauer,~H.; Sariciftci,~N.~S. Conjugated Polymer-Based
  Organic Solar Cells. \emph{Chemical Reviews} \textbf{2007}, \emph{107},
  1324--1338\relax
\mciteBstWouldAddEndPuncttrue
\mciteSetBstMidEndSepPunct{\mcitedefaultmidpunct}
{\mcitedefaultendpunct}{\mcitedefaultseppunct}\relax
\EndOfBibitem
\bibitem[Heeger(2010)]{Heeger2010}
Heeger,~A.~J. Semiconducting polymers: the Third Generation. \emph{Chem. Soc.
  Rev.} \textbf{2010}, \emph{39}, 2354--2371\relax
\mciteBstWouldAddEndPuncttrue
\mciteSetBstMidEndSepPunct{\mcitedefaultmidpunct}
{\mcitedefaultendpunct}{\mcitedefaultseppunct}\relax
\EndOfBibitem
\bibitem[Facchetti(2011)]{Facchetti2011}
Facchetti,~A. $\pi$-Conjugated Polymers for Organic Electronics and
  Photovoltaic Cell Applications. \emph{Chemistry of Materials} \textbf{2011},
  \emph{23}, 733--758\relax
\mciteBstWouldAddEndPuncttrue
\mciteSetBstMidEndSepPunct{\mcitedefaultmidpunct}
{\mcitedefaultendpunct}{\mcitedefaultseppunct}\relax
\EndOfBibitem
\bibitem[Wang \latin{et~al.}(2012)Wang, Dong, Hu, Liu, and Zhu]{Wang2012}
Wang,~C.; Dong,~H.; Hu,~W.; Liu,~Y.; Zhu,~D. Semiconducting $\pi$-Conjugated
  Systems in Field-Effect Transistors: A Material Odyssey of Organic
  Electronics. \emph{Chemical Reviews} \textbf{2012}, \emph{112},
  2208--2267\relax
\mciteBstWouldAddEndPuncttrue
\mciteSetBstMidEndSepPunct{\mcitedefaultmidpunct}
{\mcitedefaultendpunct}{\mcitedefaultseppunct}\relax
\EndOfBibitem
\bibitem[Yuen \latin{et~al.}(2011)Yuen, Fan, Seifter, Lim, Hufschmid, Heeger,
  and Wudl]{Yuen2011}
Yuen,~J.~D.; Fan,~J.; Seifter,~J.; Lim,~B.; Hufschmid,~R.; Heeger,~A.~J.;
  Wudl,~F. High Performance Weak Donor-Acceptor Polymers in Thin Film
  Transistors: Effect of the Acceptor on Electronic Properties, Ambipolar
  Conductivity, Mobility, and Thermal Stability. \emph{Journal of the American
  Chemical Society} \textbf{2011}, \emph{133}, 20799--20807\relax
\mciteBstWouldAddEndPuncttrue
\mciteSetBstMidEndSepPunct{\mcitedefaultmidpunct}
{\mcitedefaultendpunct}{\mcitedefaultseppunct}\relax
\EndOfBibitem
\bibitem[Li(2012)]{Li2012}
Li,~Y. Molecular Design of Photovoltaic Materials for Polymer Solar Cells:
  Toward Suitable Electronic Energy Levels and Broad Absorption. \emph{Accounts
  of Chemical Research} \textbf{2012}, \emph{45}, 723--733\relax
\mciteBstWouldAddEndPuncttrue
\mciteSetBstMidEndSepPunct{\mcitedefaultmidpunct}
{\mcitedefaultendpunct}{\mcitedefaultseppunct}\relax
\EndOfBibitem
\bibitem[Zhang \latin{et~al.}(2013)Zhang, Bronstein, Kronemeijer, Smith, Kim,
  Kline, Richter, Anthopoulos, Sirringhaus, Song, Heeney, Zhang, McCulloch, and
  DeLongchamp]{Zhang2013}
Zhang,~X.; Bronstein,~H.; Kronemeijer,~A.~J.; Smith,~J.; Kim,~Y.; Kline,~R.~J.;
  Richter,~L.~J.; Anthopoulos,~T.~D.; Sirringhaus,~H.; Song,~K. \latin{et~al.}
  Molecular origin of high field-effect mobility in an
  indacenodithiophene-benzothiadiazole copolymer. \emph{Nature Communications}
  \textbf{2013}, \emph{4}, 2238\relax
\mciteBstWouldAddEndPuncttrue
\mciteSetBstMidEndSepPunct{\mcitedefaultmidpunct}
{\mcitedefaultendpunct}{\mcitedefaultseppunct}\relax
\EndOfBibitem
\bibitem[Schwartz(2003)]{Schwartz2003}
Schwartz,~B.~J. Conjugated Polymers as Molecular Materials: How Chain
  Conformation and Film Morphology Influence Energy Transfer and Interchain
  Interactions. \emph{Annual Review of Physical Chemistry} \textbf{2003},
  \emph{54}, 141--172, PMID: 12524429\relax
\mciteBstWouldAddEndPuncttrue
\mciteSetBstMidEndSepPunct{\mcitedefaultmidpunct}
{\mcitedefaultendpunct}{\mcitedefaultseppunct}\relax
\EndOfBibitem
\bibitem[Noriega \latin{et~al.}(2013)Noriega, Rivnay, Vandewal, Koch,
  Stingelin, Smith, Toney, and Salleo]{Noriega2013}
Noriega,~R.; Rivnay,~J.; Vandewal,~K.; Koch,~F. P.~V.; Stingelin,~N.;
  Smith,~P.; Toney,~M.~F.; Salleo,~A. A general relationship between disorder,
  aggregation and charge transport in conjugated polymers. \emph{Nature
  Materials} \textbf{2013}, \emph{12}, 1038--1044\relax
\mciteBstWouldAddEndPuncttrue
\mciteSetBstMidEndSepPunct{\mcitedefaultmidpunct}
{\mcitedefaultendpunct}{\mcitedefaultseppunct}\relax
\EndOfBibitem
\bibitem[Oh \latin{et~al.}(2009)Oh, Lee, Mannsfeld, Stoltenberg, Jung, Jin,
  Kim, Yoo, and Bao]{Oh2009}
Oh,~J.~H.; Lee,~H.~W.; Mannsfeld,~S.; Stoltenberg,~R.~M.; Jung,~E.; Jin,~Y.~W.;
  Kim,~J.~M.; Yoo,~J.-B.; Bao,~Z. Solution-processed, high-performance
  n-channel organic microwire transistors. \emph{Proceedings of the National
  Academy of Sciences} \textbf{2009}, \emph{106}, 6065--6070\relax
\mciteBstWouldAddEndPuncttrue
\mciteSetBstMidEndSepPunct{\mcitedefaultmidpunct}
{\mcitedefaultendpunct}{\mcitedefaultseppunct}\relax
\EndOfBibitem
\bibitem[Barzegar \latin{et~al.}(2018)Barzegar, Larsen, Boulanger, Zettl,
  Edman, and W{\aa}gberg]{Barzegar2018}
Barzegar,~H.~R.; Larsen,~C.; Boulanger,~N.; Zettl,~A.; Edman,~L.;
  W{\aa}gberg,~T. Self-Assembled PCBM Nanosheets: A Facile Route to Electronic
  Layer-on-Layer Heterostructures. \emph{Nano Letters} \textbf{2018},
  \emph{18}, 1442--1447\relax
\mciteBstWouldAddEndPuncttrue
\mciteSetBstMidEndSepPunct{\mcitedefaultmidpunct}
{\mcitedefaultendpunct}{\mcitedefaultseppunct}\relax
\EndOfBibitem
\bibitem[Sirringhaus \latin{et~al.}(1999)Sirringhaus, Brown, Friend, Nielsen,
  Bechgaard, Langeveld-Voss, Spiering, Janssen, Meijer, Herwig, and
  de~Leeuw]{Sirringhaus1999}
Sirringhaus,~H.; Brown,~P.~J.; Friend,~R.~H.; Nielsen,~M.~M.; Bechgaard,~K.;
  Langeveld-Voss,~B. M.~W.; Spiering,~A. J.~H.; Janssen,~R. A.~J.;
  Meijer,~E.~W.; Herwig,~P. \latin{et~al.}  Two-dimensional charge transport in
  self-organized, high-mobility conjugated polymers. \emph{Nature}
  \textbf{1999}, \emph{401}, 685--688\relax
\mciteBstWouldAddEndPuncttrue
\mciteSetBstMidEndSepPunct{\mcitedefaultmidpunct}
{\mcitedefaultendpunct}{\mcitedefaultseppunct}\relax
\EndOfBibitem
\bibitem[Venkateshvaran \latin{et~al.}(2014)Venkateshvaran, Nikolka, Sadhanala,
  Lemaur, Zelazny, Kepa, Hurhangee, Kronemeijer, Pecunia, Nasrallah, Romanov,
  Broch, McCulloch, Emin, Olivier, Cornil, Beljonne, and
  Sirringhaus]{Venkateshvaran2014}
Venkateshvaran,~D.; Nikolka,~M.; Sadhanala,~A.; Lemaur,~V.; Zelazny,~M.;
  Kepa,~M.; Hurhangee,~M.; Kronemeijer,~A.~J.; Pecunia,~V.; Nasrallah,~I.
  \latin{et~al.}  Approaching disorder-free transport in high-mobility
  conjugated polymers. \emph{Nature} \textbf{2014}, \emph{515}, 384--388\relax
\mciteBstWouldAddEndPuncttrue
\mciteSetBstMidEndSepPunct{\mcitedefaultmidpunct}
{\mcitedefaultendpunct}{\mcitedefaultseppunct}\relax
\EndOfBibitem
\bibitem[Luo \latin{et~al.}(2014)Luo, Kyaw, Perez, Patel, Wang, Grimm, Bazan,
  Kramer, and Heeger]{Luo2014}
Luo,~C.; Kyaw,~A. K.~K.; Perez,~L.~A.; Patel,~S.; Wang,~M.; Grimm,~B.;
  Bazan,~G.~C.; Kramer,~E.~J.; Heeger,~A.~J. General Strategy for Self-Assembly
  of Highly Oriented Nanocrystalline Semiconducting Polymers with High
  Mobility. \emph{Nano Letters} \textbf{2014}, \emph{14}, 2764--2771\relax
\mciteBstWouldAddEndPuncttrue
\mciteSetBstMidEndSepPunct{\mcitedefaultmidpunct}
{\mcitedefaultendpunct}{\mcitedefaultseppunct}\relax
\EndOfBibitem
\bibitem[Li \latin{et~al.}(2016)Li, An, Marszalek, Baumgarten, Yan, M{\"u}llen,
  and Pisula]{Li2016}
Li,~M.; An,~C.; Marszalek,~T.; Baumgarten,~M.; Yan,~H.; M{\"u}llen,~K.;
  Pisula,~W. Controlling the Surface Organization of Conjugated Donor-Acceptor
  Polymers by their Aggregation in Solution. \emph{Advanced Materials}
  \textbf{2016}, \emph{28}, 9430--9438\relax
\mciteBstWouldAddEndPuncttrue
\mciteSetBstMidEndSepPunct{\mcitedefaultmidpunct}
{\mcitedefaultendpunct}{\mcitedefaultseppunct}\relax
\EndOfBibitem
\bibitem[Hsu \latin{et~al.}(2015)Hsu, Cheng, Luo, Patel, Zhong, Sun, Sherman,
  Lee, Ying, Wang, Bazan, Chabinyc, Br{\'e}das, and Heeger]{Hsu2015}
Hsu,~B. B.-Y.; Cheng,~C.-M.; Luo,~C.; Patel,~S.~N.; Zhong,~C.; Sun,~H.;
  Sherman,~J.; Lee,~B.~H.; Ying,~L.; Wang,~M. \latin{et~al.}  The Density of
  States and the Transport Effective Mass in a Highly Oriented Semiconducting
  Polymer: Electronic Delocalization in 1D. \emph{Advanced Materials}
  \textbf{2015}, \emph{27}, 7759--7765\relax
\mciteBstWouldAddEndPuncttrue
\mciteSetBstMidEndSepPunct{\mcitedefaultmidpunct}
{\mcitedefaultendpunct}{\mcitedefaultseppunct}\relax
\EndOfBibitem
\bibitem[Sakanoue and Sirringhaus(2010)Sakanoue, and Sirringhaus]{Sakanoue2010}
Sakanoue,~T.; Sirringhaus,~H. Band-like temperature dependence of mobility in a
  solution-processed organic semiconductor. \emph{Nature Materials}
  \textbf{2010}, \emph{9}, 736--740\relax
\mciteBstWouldAddEndPuncttrue
\mciteSetBstMidEndSepPunct{\mcitedefaultmidpunct}
{\mcitedefaultendpunct}{\mcitedefaultseppunct}\relax
\EndOfBibitem
\bibitem[Yamashita \latin{et~al.}(2014)Yamashita, Tsurumi, Hinkel, Okada,
  Soeda, Zajaczkowski, Baumgarten, Pisula, Matsui, M{\"u}llen, and
  Takeya]{Yamashita2014}
Yamashita,~Y.; Tsurumi,~J.; Hinkel,~F.; Okada,~Y.; Soeda,~J.; Zajaczkowski,~W.;
  Baumgarten,~M.; Pisula,~W.; Matsui,~H.; M{\"u}llen,~K. \latin{et~al.}
  Transition Between Band and Hopping Transport in Polymer Field-Effect
  Transistors. \emph{Advanced Materials} \textbf{2014}, \emph{26},
  8169--8173\relax
\mciteBstWouldAddEndPuncttrue
\mciteSetBstMidEndSepPunct{\mcitedefaultmidpunct}
{\mcitedefaultendpunct}{\mcitedefaultseppunct}\relax
\EndOfBibitem
\bibitem[Kim \latin{et~al.}(2014)Kim, Kang, Dutta, Han, Shin, Noh, and
  Yang]{Kim2014}
Kim,~G.; Kang,~S.-J.; Dutta,~G.~K.; Han,~Y.-K.; Shin,~T.~J.; Noh,~Y.-Y.;
  Yang,~C. A Thienoisoindigo-Naphthalene Polymer with Ultrahigh Mobility of
  14.4 cm2/V{\textperiodcentered}s That Substantially Exceeds Benchmark Values
  for Amorphous Silicon Semiconductors. \emph{Journal of the American Chemical
  Society} \textbf{2014}, \emph{136}, 9477--9483\relax
\mciteBstWouldAddEndPuncttrue
\mciteSetBstMidEndSepPunct{\mcitedefaultmidpunct}
{\mcitedefaultendpunct}{\mcitedefaultseppunct}\relax
\EndOfBibitem
\bibitem[Sutton \latin{et~al.}(2016)Sutton, Risko, and Br{\'e}das]{Sutton2016}
Sutton,~C.; Risko,~C.; Br{\'e}das,~J.-L. Noncovalent Intermolecular
  Interactions in Organic Electronic Materials: Implications for the Molecular
  Packing vs Electronic Properties of Acenes. \emph{Chemistry of Materials}
  \textbf{2016}, \emph{28}, 3--16\relax
\mciteBstWouldAddEndPuncttrue
\mciteSetBstMidEndSepPunct{\mcitedefaultmidpunct}
{\mcitedefaultendpunct}{\mcitedefaultseppunct}\relax
\EndOfBibitem
\bibitem[Norton and Br{\'e}das(2008)Norton, and Br{\'e}das]{Norton2008}
Norton,~J.~E.; Br{\'e}das,~J.-L. Polarization Energies in Oligoacene
  Semiconductor Crystals. \emph{Journal of the American Chemical Society}
  \textbf{2008}, \emph{130}, 12377--12384\relax
\mciteBstWouldAddEndPuncttrue
\mciteSetBstMidEndSepPunct{\mcitedefaultmidpunct}
{\mcitedefaultendpunct}{\mcitedefaultseppunct}\relax
\EndOfBibitem
\bibitem[Nayak and Periasamy(2009)Nayak, and Periasamy]{Nayak2009}
Nayak,~P.~K.; Periasamy,~N. Calculation of electron affinity, ionization
  potential, transport gap, optical band gap and exciton binding energy of
  organic solids using ‘solvation’ model and DFT. \emph{Organic
  Electronics} \textbf{2009}, \emph{10}, 1396--1400\relax
\mciteBstWouldAddEndPuncttrue
\mciteSetBstMidEndSepPunct{\mcitedefaultmidpunct}
{\mcitedefaultendpunct}{\mcitedefaultseppunct}\relax
\EndOfBibitem
\bibitem[Difley \latin{et~al.}(2010)Difley, Wang, Yeganeh, Yost, and
  Voorhis]{Difley2010}
Difley,~S.; Wang,~L.-P.; Yeganeh,~S.; Yost,~S.~R.; Voorhis,~T.~V. Electronic
  Properties of Disordered Organic Semiconductors via QM/MM Simulations.
  \emph{Accounts of Chemical Research} \textbf{2010}, \emph{43},
  995--1004\relax
\mciteBstWouldAddEndPuncttrue
\mciteSetBstMidEndSepPunct{\mcitedefaultmidpunct}
{\mcitedefaultendpunct}{\mcitedefaultseppunct}\relax
\EndOfBibitem
\bibitem[Ryno \latin{et~al.}(2013)Ryno, Lee, Sears, Risko, and
  Br{\'e}das]{Ryno2013}
Ryno,~S.~M.; Lee,~S.~R.; Sears,~J.~S.; Risko,~C.; Br{\'e}das,~J.-L. Electronic
  Polarization Effects upon Charge Injection in Oligoacene Molecular Crystals:
  Description via a Polarizable Force Field. \emph{The Journal of Physical
  Chemistry C} \textbf{2013}, \emph{117}, 13853--13860\relax
\mciteBstWouldAddEndPuncttrue
\mciteSetBstMidEndSepPunct{\mcitedefaultmidpunct}
{\mcitedefaultendpunct}{\mcitedefaultseppunct}\relax
\EndOfBibitem
\bibitem[Refaely-Abramson \latin{et~al.}(2013)Refaely-Abramson, Sharifzadeh,
  Jain, Baer, Neaton, and Kronik]{Refaely-Abramson2013}
Refaely-Abramson,~S.; Sharifzadeh,~S.; Jain,~M.; Baer,~R.; Neaton,~J.~B.;
  Kronik,~L. Gap renormalization of molecular crystals from density-functional
  theory. \emph{Physical Review B} \textbf{2013}, \emph{88}, 081204\relax
\mciteBstWouldAddEndPuncttrue
\mciteSetBstMidEndSepPunct{\mcitedefaultmidpunct}
{\mcitedefaultendpunct}{\mcitedefaultseppunct}\relax
\EndOfBibitem
\bibitem[Poelking \latin{et~al.}(2015)Poelking, Tietze, Elschner, Olthof,
  Hertel, Baumeier, W{\"u}rthner, Meerholz, Leo, and Andrienko]{Poelking2015}
Poelking,~C.; Tietze,~M.; Elschner,~C.; Olthof,~S.; Hertel,~D.; Baumeier,~B.;
  W{\"u}rthner,~F.; Meerholz,~K.; Leo,~K.; Andrienko,~D. Impact of mesoscale
  order on open-circuit voltage in organic solar cells. \emph{Nature Materials}
  \textbf{2015}, \emph{14}, 434--439\relax
\mciteBstWouldAddEndPuncttrue
\mciteSetBstMidEndSepPunct{\mcitedefaultmidpunct}
{\mcitedefaultendpunct}{\mcitedefaultseppunct}\relax
\EndOfBibitem
\bibitem[Kang \latin{et~al.}(2016)Kang, Jeon, Cho, and Han]{Kang2016}
Kang,~Y.; Jeon,~S.~H.; Cho,~Y.; Han,~S. Ab initio calculation of ionization
  potential and electron affinity in solid-state organic semiconductors.
  \emph{Phys. Rev. B} \textbf{2016}, \emph{93}, 035131\relax
\mciteBstWouldAddEndPuncttrue
\mciteSetBstMidEndSepPunct{\mcitedefaultmidpunct}
{\mcitedefaultendpunct}{\mcitedefaultseppunct}\relax
\EndOfBibitem
\bibitem[Li \latin{et~al.}(2016)Li, D'Avino, Duchemin, Beljonne, and
  Blase]{Li2016_Blase}
Li,~J.; D'Avino,~G.; Duchemin,~I.; Beljonne,~D.; Blase,~X. Combining the
  Many-Body GW Formalism with Classical Polarizable Models: Insights on the
  Electronic Structure of Molecular Solids. \emph{The Journal of Physical
  Chemistry Letters} \textbf{2016}, \emph{7}, 2814--2820\relax
\mciteBstWouldAddEndPuncttrue
\mciteSetBstMidEndSepPunct{\mcitedefaultmidpunct}
{\mcitedefaultendpunct}{\mcitedefaultseppunct}\relax
\EndOfBibitem
\bibitem[Sun \latin{et~al.}(2016)Sun, Ryno, Zhong, Ravva, Sun,
  K{\"o}rzd{\"o}rfer, and Br{\'e}das]{Sun2016}
Sun,~H.; Ryno,~S.; Zhong,~C.; Ravva,~M.~K.; Sun,~Z.; K{\"o}rzd{\"o}rfer,~T.;
  Br{\'e}das,~J.-L. Ionization Energies, Electron Affinities, and Polarization
  Energies of Organic Molecular Crystals: Quantitative Estimations from a
  Polarizable Continuum Model (PCM)-Tuned Range-Separated Density Functional
  Approach. \emph{Journal of Chemical Theory and Computation} \textbf{2016},
  \emph{12}, 2906--2916\relax
\mciteBstWouldAddEndPuncttrue
\mciteSetBstMidEndSepPunct{\mcitedefaultmidpunct}
{\mcitedefaultendpunct}{\mcitedefaultseppunct}\relax
\EndOfBibitem
\bibitem[Li \latin{et~al.}(2018)Li, D'Avino, Duchemin, Beljonne, and
  Blase]{Li2018}
Li,~J.; D'Avino,~G.; Duchemin,~I.; Beljonne,~D.; Blase,~X. Accurate description
  of charged excitations in molecular solids from embedded many-body
  perturbation theory. \emph{Phys. Rev. B} \textbf{2018}, \emph{97},
  035108\relax
\mciteBstWouldAddEndPuncttrue
\mciteSetBstMidEndSepPunct{\mcitedefaultmidpunct}
{\mcitedefaultendpunct}{\mcitedefaultseppunct}\relax
\EndOfBibitem
\bibitem[Bhandari \latin{et~al.}(2018)Bhandari, Cheung, Geva, Kronik, and
  Dunietz]{Bhandari2018}
Bhandari,~S.; Cheung,~M.~S.; Geva,~E.; Kronik,~L.; Dunietz,~B.~D. Fundamental
  Gaps of Condensed-Phase Organic Semiconductors from Single-Molecule
  Calculations using Polarization-Consistent Optimally Tuned Screened
  Range-Separated Hybrid Functionals. \emph{Journal of Chemical Theory and
  Computation} \textbf{2018}, \emph{14}, 6287--6294\relax
\mciteBstWouldAddEndPuncttrue
\mciteSetBstMidEndSepPunct{\mcitedefaultmidpunct}
{\mcitedefaultendpunct}{\mcitedefaultseppunct}\relax
\EndOfBibitem
\bibitem[Halls \latin{et~al.}(1999)Halls, Cornil, dos Santos, Silbey, Hwang,
  Holmes, Br\'edas, and Friend]{Halls1999}
Halls,~J. J.~M.; Cornil,~J.; dos Santos,~D.~A.; Silbey,~R.; Hwang,~D.-H.;
  Holmes,~A.~B.; Br\'edas,~J.~L.; Friend,~R.~H. Charge- and energy-transfer
  processes at polymer/polymer interfaces: A joint experimental and theoretical
  study. \emph{Phys. Rev. B} \textbf{1999}, \emph{60}, 5721--5727\relax
\mciteBstWouldAddEndPuncttrue
\mciteSetBstMidEndSepPunct{\mcitedefaultmidpunct}
{\mcitedefaultendpunct}{\mcitedefaultseppunct}\relax
\EndOfBibitem
\bibitem[Cornil \latin{et~al.}(2003)Cornil, Gueli, Dkhissi, Sancho-Garcia,
  Hennebicq, Calbert, Lemaur, Beljonne, and Br{\'e}das]{Cornil2003}
Cornil,~J.; Gueli,~I.; Dkhissi,~A.; Sancho-Garcia,~J.~C.; Hennebicq,~E.;
  Calbert,~J.~P.; Lemaur,~V.; Beljonne,~D.; Br{\'e}das,~J.~L. Electronic and
  optical properties of polyfluorene and fluorene-based copolymers: A
  quantum-chemical characterization. \emph{The Journal of Chemical Physics}
  \textbf{2003}, \emph{118}, 6615--6623\relax
\mciteBstWouldAddEndPuncttrue
\mciteSetBstMidEndSepPunct{\mcitedefaultmidpunct}
{\mcitedefaultendpunct}{\mcitedefaultseppunct}\relax
\EndOfBibitem
\bibitem[Br\'edas \latin{et~al.}(1984)Br\'edas, Th\'emans, Fripiat, Andr\'e,
  and Chance]{Bredas1984}
Br\'edas,~J.~L.; Th\'emans,~B.; Fripiat,~J.~G.; Andr\'e,~J.~M.; Chance,~R.~R.
  Highly conducting polyparaphenylene, polypyrrole, and polythiophene chains:
  An ab initio study of the geometry and electronic-structure modifications
  upon doping. \emph{Phys. Rev. B} \textbf{1984}, \emph{29}, 6761--6773\relax
\mciteBstWouldAddEndPuncttrue
\mciteSetBstMidEndSepPunct{\mcitedefaultmidpunct}
{\mcitedefaultendpunct}{\mcitedefaultseppunct}\relax
\EndOfBibitem
\bibitem[Br{\'e}das \latin{et~al.}(1985)Br{\'e}das, Street, Th{\'e}mans, and
  Andr{\'e}]{Bredas1985}
Br{\'e}das,~J.~L.; Street,~G.~B.; Th{\'e}mans,~B.; Andr{\'e},~J.~M. Organic
  polymers based on aromatic rings (polyparaphenylene, polypyrrole,
  polythiophene): Evolution of the electronic properties as a function of the
  torsion angle between adjacent rings. \emph{The Journal of Chemical Physics}
  \textbf{1985}, \emph{83}, 1323--1329\relax
\mciteBstWouldAddEndPuncttrue
\mciteSetBstMidEndSepPunct{\mcitedefaultmidpunct}
{\mcitedefaultendpunct}{\mcitedefaultseppunct}\relax
\EndOfBibitem
\bibitem[Cheng \latin{et~al.}(2017)Cheng, Geng, Yi, and Shuai]{cheng2017}
Cheng,~C.; Geng,~H.; Yi,~Y.; Shuai,~Z. Super-exchange-induced high performance
  charge transport in donor-acceptor copolymers. \emph{Journal of Materials
  Chemistry C} \textbf{2017}, \emph{5}, 3247--3253\relax
\mciteBstWouldAddEndPuncttrue
\mciteSetBstMidEndSepPunct{\mcitedefaultmidpunct}
{\mcitedefaultendpunct}{\mcitedefaultseppunct}\relax
\EndOfBibitem
\bibitem[He \latin{et~al.}(2018)He, Cheng, Geng, Yi, and Shuai]{He2018}
He,~F.; Cheng,~C.; Geng,~H.; Yi,~Y.; Shuai,~Z. Effect of donor length on
  electronic structures and charge transport polarity for DTDPP-based D-A
  copolymers: a computational study based on a super-exchange model.
  \emph{Journal of Materials Chemistry A} \textbf{2018}, \emph{6},
  11985--11993\relax
\mciteBstWouldAddEndPuncttrue
\mciteSetBstMidEndSepPunct{\mcitedefaultmidpunct}
{\mcitedefaultendpunct}{\mcitedefaultseppunct}\relax
\EndOfBibitem
\bibitem[Br{\'e}das \latin{et~al.}(2018)Br{\'e}das, Li, Sun, and
  Zhong]{Bredas2018}
Br{\'e}das,~J.-L.; Li,~Y.; Sun,~H.; Zhong,~C. Why Can High Charge-Carrier
  Mobilities be Achieved Along p-Conjugated Polymer Chains with Alternating
  Donor-Acceptor Moieties? \emph{Advanced Theory and Simulations}
  \textbf{2018}, \emph{1}, 1800016\relax
\mciteBstWouldAddEndPuncttrue
\mciteSetBstMidEndSepPunct{\mcitedefaultmidpunct}
{\mcitedefaultendpunct}{\mcitedefaultseppunct}\relax
\EndOfBibitem
\bibitem[Woods \latin{et~al.}(2016)Woods, Dalvit, Tkatchenko, Rodriguez-Lopez,
  Rodriguez, and Podgornik]{Woods2016}
Woods,~L.~M.; Dalvit,~D. A.~R.; Tkatchenko,~A.; Rodriguez-Lopez,~P.;
  Rodriguez,~A.~W.; Podgornik,~R. Materials perspective on Casimir and van der
  Waals interactions. \emph{Rev. Mod. Phys.} \textbf{2016}, \emph{88},
  045003\relax
\mciteBstWouldAddEndPuncttrue
\mciteSetBstMidEndSepPunct{\mcitedefaultmidpunct}
{\mcitedefaultendpunct}{\mcitedefaultseppunct}\relax
\EndOfBibitem
\bibitem[Martin \latin{et~al.}(2016)Martin, Reining, and
  Ceperley]{martin_reining_ceperley_2016}
Martin,~R.~M.; Reining,~L.; Ceperley,~D.~M. \emph{Interacting Electrons: Theory
  and Computational Approaches}; Cambridge University Press, 2016\relax
\mciteBstWouldAddEndPuncttrue
\mciteSetBstMidEndSepPunct{\mcitedefaultmidpunct}
{\mcitedefaultendpunct}{\mcitedefaultseppunct}\relax
\EndOfBibitem
\bibitem[Neuhauser \latin{et~al.}(2014)Neuhauser, Gao, Arntsen, Karshenas,
  Rabani, and Baer]{Neuhauser2014}
Neuhauser,~D.; Gao,~Y.; Arntsen,~C.; Karshenas,~C.; Rabani,~E.; Baer,~R.
  Breaking the Theoretical Scaling Limit for Predicting Quasiparticle Energies:
  The Stochastic $GW$ Approach. \emph{Phys. Rev. Lett.} \textbf{2014},
  \emph{113}, 076402\relax
\mciteBstWouldAddEndPuncttrue
\mciteSetBstMidEndSepPunct{\mcitedefaultmidpunct}
{\mcitedefaultendpunct}{\mcitedefaultseppunct}\relax
\EndOfBibitem
\bibitem[Vl\v{c}ek \latin{et~al.}(2017)Vl\v{c}ek, Rabani, Neuhauser, and
  Baer]{Vlcek2017}
Vl\v{c}ek,~V.; Rabani,~E.; Neuhauser,~D.; Baer,~R. Stochastic GW Calculations
  for Molecules. \emph{Journal of Chemical Theory and Computation}
  \textbf{2017}, \emph{13}, 4997--5003, PMID: 28876912\relax
\mciteBstWouldAddEndPuncttrue
\mciteSetBstMidEndSepPunct{\mcitedefaultmidpunct}
{\mcitedefaultendpunct}{\mcitedefaultseppunct}\relax
\EndOfBibitem
\bibitem[Vl\v{c}ek \latin{et~al.}(2018)Vl\v{c}ek, Li, Baer, Rabani, and
  Neuhauser]{Vlcek2018}
Vl\v{c}ek,~V.; Li,~W.; Baer,~R.; Rabani,~E.; Neuhauser,~D. Swift $GW$ beyond
  10,000 electrons using sparse stochastic compression. \emph{Phys. Rev. B}
  \textbf{2018}, \emph{98}, 075107\relax
\mciteBstWouldAddEndPuncttrue
\mciteSetBstMidEndSepPunct{\mcitedefaultmidpunct}
{\mcitedefaultendpunct}{\mcitedefaultseppunct}\relax
\EndOfBibitem
\bibitem[Vl\v{c}ek(2019)]{Vlcek2019}
Vl\v{c}ek,~V. Stochastic Vertex Corrections: Linear Scaling Methods for
  Accurate Quasiparticle Energies. \emph{Journal of Chemical Theory and
  Computation} \textbf{2019}, \emph{15}, 6254--6266\relax
\mciteBstWouldAddEndPuncttrue
\mciteSetBstMidEndSepPunct{\mcitedefaultmidpunct}
{\mcitedefaultendpunct}{\mcitedefaultseppunct}\relax
\EndOfBibitem
\bibitem[Blase \latin{et~al.}(2011)Blase, Attaccalite, and Olevano]{Blase2011}
Blase,~X.; Attaccalite,~C.; Olevano,~V. First-principles $\mathit{GW}$
  calculations for fullerenes, porphyrins, phtalocyanine, and other molecules
  of interest for organic photovoltaic applications. \emph{Phys. Rev. B}
  \textbf{2011}, \emph{83}, 115103\relax
\mciteBstWouldAddEndPuncttrue
\mciteSetBstMidEndSepPunct{\mcitedefaultmidpunct}
{\mcitedefaultendpunct}{\mcitedefaultseppunct}\relax
\EndOfBibitem
\bibitem[Mai \latin{et~al.}(2013)Mai, Zhou, Zhang, Henson, Nguyen, Heeger, and
  Bazan]{Mai2013}
Mai,~C.-K.; Zhou,~H.; Zhang,~Y.; Henson,~Z.~B.; Nguyen,~T.-Q.; Heeger,~A.~J.;
  Bazan,~G.~C. Facile Doping of Anionic Narrow-Band-Gap Conjugated
  Polyelectrolytes During Dialysis. \emph{Angewandte Chemie International
  Edition} \textbf{2013}, \emph{52}, 12874--12878\relax
\mciteBstWouldAddEndPuncttrue
\mciteSetBstMidEndSepPunct{\mcitedefaultmidpunct}
{\mcitedefaultendpunct}{\mcitedefaultseppunct}\relax
\EndOfBibitem
\bibitem[Mai \latin{et~al.}(2015)Mai, Russ, Fronk, Hu, Chan-Park, Urban,
  Segalman, Chabinyc, and Bazan]{Mai2015}
Mai,~C.-K.; Russ,~B.; Fronk,~S.~L.; Hu,~N.; Chan-Park,~M.~B.; Urban,~J.~J.;
  Segalman,~R.~A.; Chabinyc,~M.~L.; Bazan,~G.~C. Varying the ionic
  functionalities of conjugated polyelectrolytes leads to both p- and n-type
  carbon nanotube composites for flexible thermoelectrics. \emph{Energy \&
  Environmental Science} \textbf{2015}, \emph{8}, 2341--2346\relax
\mciteBstWouldAddEndPuncttrue
\mciteSetBstMidEndSepPunct{\mcitedefaultmidpunct}
{\mcitedefaultendpunct}{\mcitedefaultseppunct}\relax
\EndOfBibitem
\bibitem[Cui and Bazan(2018)Cui, and Bazan]{Cui2018}
Cui,~Q.; Bazan,~G.~C. Narrow Band Gap Conjugated Polyelectrolytes.
  \emph{Accounts of Chemical Research} \textbf{2018}, \emph{51}, 202--211\relax
\mciteBstWouldAddEndPuncttrue
\mciteSetBstMidEndSepPunct{\mcitedefaultmidpunct}
{\mcitedefaultendpunct}{\mcitedefaultseppunct}\relax
\EndOfBibitem
\bibitem[Zhou \latin{et~al.}(2012)Zhou, Yang, and You]{Zhou2012}
Zhou,~H.; Yang,~L.; You,~W. Rational Design of High Performance Conjugated
  Polymers for Organic Solar Cells. \emph{Macromolecules} \textbf{2012},
  \emph{45}, 607--632\relax
\mciteBstWouldAddEndPuncttrue
\mciteSetBstMidEndSepPunct{\mcitedefaultmidpunct}
{\mcitedefaultendpunct}{\mcitedefaultseppunct}\relax
\EndOfBibitem
\bibitem[Duan \latin{et~al.}(2012)Duan, Huang, and Cao]{Duan2012}
Duan,~C.; Huang,~F.; Cao,~Y. Recent development of push-pull conjugated
  polymers for bulk-heterojunction photovoltaics: rational design and fine
  tailoring of molecular structures. \emph{Journal of Materials Chemistry}
  \textbf{2012}, \emph{22}, 10416--10434\relax
\mciteBstWouldAddEndPuncttrue
\mciteSetBstMidEndSepPunct{\mcitedefaultmidpunct}
{\mcitedefaultendpunct}{\mcitedefaultseppunct}\relax
\EndOfBibitem
\bibitem[one()]{onebodyterm}
The one-body terms include the single particle non-interacting kinetic energy
  and external as well as the Hartree potential energies.\relax
\mciteBstWouldAddEndPunctfalse
\mciteSetBstMidEndSepPunct{\mcitedefaultmidpunct}
{}{\mcitedefaultseppunct}\relax
\EndOfBibitem
\bibitem[loc()]{localproperties}
The external and Hartree potentials are manifestedly local; the non-interacting
  kinetic energy is determined purely by a Kohn-Sham map using a local
  potential. Note that the many-body calculations use a perturbative correction
  with the Kohn-Sham results as a starting point. The non-local contributions
  are included in the self-energy term, i.e., it is part of the
  electron-electron interaction.\relax
\mciteBstWouldAddEndPunctfalse
\mciteSetBstMidEndSepPunct{\mcitedefaultmidpunct}
{}{\mcitedefaultseppunct}\relax
\EndOfBibitem
\bibitem[Mori-S\'anchez \latin{et~al.}(2008)Mori-S\'anchez, Cohen, and
  Yang]{Sanchez2008}
Mori-S\'anchez,~P.; Cohen,~A.~J.; Yang,~W. Localization and Delocalization
  Errors in Density Functional Theory and Implications for Band-Gap Prediction.
  \emph{Phys. Rev. Lett.} \textbf{2008}, \emph{100}, 146401\relax
\mciteBstWouldAddEndPuncttrue
\mciteSetBstMidEndSepPunct{\mcitedefaultmidpunct}
{\mcitedefaultendpunct}{\mcitedefaultseppunct}\relax
\EndOfBibitem
\bibitem[Vl\v{c}ek \latin{et~al.}(2016)Vl\v{c}ek, Eisenberg, Steinle-Neumann,
  Neuhauser, Rabani, and Baer]{Vlcek2016}
Vl\v{c}ek,~V.; Eisenberg,~H.~R.; Steinle-Neumann,~G.; Neuhauser,~D.;
  Rabani,~E.; Baer,~R. Spontaneous Charge Carrier Localization in Extended
  One-Dimensional Systems. \emph{Phys. Rev. Lett.} \textbf{2016}, \emph{116},
  186401\relax
\mciteBstWouldAddEndPuncttrue
\mciteSetBstMidEndSepPunct{\mcitedefaultmidpunct}
{\mcitedefaultendpunct}{\mcitedefaultseppunct}\relax
\EndOfBibitem
\bibitem[dis()]{dispersion}
For comparison, this value is practically identical to the dispersion of the
  conjugated bands along the polymer chain in the 2D system (c.f.,
  Figure~\ref{Fig_2DExP})\relax
\mciteBstWouldAddEndPuncttrue
\mciteSetBstMidEndSepPunct{\mcitedefaultmidpunct}
{\mcitedefaultendpunct}{\mcitedefaultseppunct}\relax
\EndOfBibitem
\bibitem[Huang \latin{et~al.}(2016)Huang, Markwart, Briseno, and
  Hayward]{Huang2016}
Huang,~W.; Markwart,~J.~C.; Briseno,~A.~L.; Hayward,~R.~C. Orthogonal Ambipolar
  Semiconductor Nanostructures for Complementary Logic Gates. \emph{ACS Nano}
  \textbf{2016}, \emph{10}, 8610--8619\relax
\mciteBstWouldAddEndPuncttrue
\mciteSetBstMidEndSepPunct{\mcitedefaultmidpunct}
{\mcitedefaultendpunct}{\mcitedefaultseppunct}\relax
\EndOfBibitem
\bibitem[Huang and Hayward(2018)Huang, and Hayward]{Huang2018}
Huang,~W.; Hayward,~R.~C. Orthogonal Ambipolar Semiconductors with Inherently
  Multi-Dimensional Responses for the Discriminative Sensing of Chemical
  Vapors. \emph{ACS Applied Materials {\&} Interfaces} \textbf{2018},
  \emph{10}, 33353--33359\relax
\mciteBstWouldAddEndPuncttrue
\mciteSetBstMidEndSepPunct{\mcitedefaultmidpunct}
{\mcitedefaultendpunct}{\mcitedefaultseppunct}\relax
\EndOfBibitem
\end{mcitethebibliography}

\end{document}


\maketitle

\section{Computational details}

\subsection{Geometries}

Each polymer is considered to be straight and infinitely periodic. For D-A copolymers, the conventional alkyl groups attached to the fluorene unit are replaced with hydrogen atoms for computational convenience. The geometry and lattice constans of each periodic system are fully relaxed by employing Quantum-Espresso package \cite{Giannozzi2009,Giannozzi2017} with Kohn-Sham density functional theory (DFT) within the generalized gradient approximation (GGA) \cite{Perdew1996} combined with Tkachenko-Scheffler treatment of the van der Waals interactions.\cite{Tkatchenko2009} The lattice parameters of rectangular cells used throughout this study are summarized in Table~\ref{tab:lattice_constants_and_k-point} together with the k--point meshes used to sample the Brillouin zones. In the optimization, $\pi$-$\pi$ stacking without displacement is energetically the most favorable; the structures are illustrated in Figure \ref{Fig_supercells}.

\begin{table}[H]
     \centering
     \begin{tabular}{c|c|c|c|ccc|c}
 \multirow{2}{*}{system} & \multicolumn{3}{c|}{lattice constants}                                                   & \multirow{2}{*}{\begin{tabular}[c]{@{}c@{}}k-point\\ mesh\end{tabular}} \\ \cline{2-4}
                         & a [\AA]     & b [\AA]    & c [\AA]             &                                                                         \\ \hline
 1D FBT                  & 12.774 & -     & -     & 4$\times$1$\times$1                                                                   \\
 2D FBT                  & 12.759 & 3.836 & -    & 1$\times$4$\times$1                                                                   \\
 3D FBT                  & 12.759 & 3.802 & 7.109   & 4$\times$4$\times$4                                                                   \\
 1D FB-Ox                 & 12.774 & -     & -      & 4$\times$1$\times$1                                                                   \\
 1D FB-Se                 & 12.774 & -     & -       & 4$\times$1$\times$1                                                                  
 \end{tabular}
     \caption{Lattice constants of various systems of interest and the k-point meshes for geometry optimization used in the Quantum-Espresso package.}
     \label{tab:lattice_constants_and_k-point}
 \end{table}
 
\subsection{Electronic structure and excitation energies}
The many-body calculations are performed on polymer geometries obtained from first-principles structural optimizations with 0D, 1D, 2D, and 3D periodic boundary conditions.\cite{Rozzi2006} The details of the structural relaxation are provided the previous section. The QP energies are obtained from perturbation theory using the Kohn-Sham density functional theory (DFT) with generalized gradient approximation as a starting point. The DFT step was performed using supercells that correspond to the $k$-point meshes in Table~\ref{tab:lattice_constants_and_k-point}; the supercells are illustrated in Figure~\ref{Fig_supercells}. The parameters of our DFT calculations are in Table~\ref{tab:DFTsetups1}\&\ref{tab:DFTsetups2}.
From the DFT step, we obtain a set of eigenvalues $\left\{\varepsilon^{\rm KS}\right\}$ and corresponding eigenstates $\left\{\phi\right\}$. The quasiparticle energies are computed as
\begin{equation}
    \varepsilon = \varepsilon^{\rm KS} + \left\langle \phi\middle| \hat \Sigma(\omega = \varepsilon) - \hat v_{xc} \middle| \phi\right\rangle
\end{equation}
where $v_{xc}$ is the mean-field exchange-correlation potential and $\Sigma(\omega)$ is the dynamical and non-local self-energy operator. In the space-time domain (represented by coordinates $1\equiv (r_1, t_1)$, the self-energy is approximated as \cite{Hedin1965}
\begin{equation}
    \Sigma(1,2) = iG_0(1,2)W_0(1,2^+)
\end{equation}
where $G$ is the KS Green's function, $W$ is the screened Coulomb interaction computed within the random-phase approximation, and $2^+$ is infinitesimally later than $2$.\cite{martin_reining_ceperley_2016} The total self-energy is further decomposed to the static exchange and frequency-dependent correlation contribution. The evaluation of the self-energy employs the stochastic approach in which the expectation value of the self-energy is computed through a randomized sampling of wave functions and stochastic decomposition of quantum mechanical operators.\cite{Neuhauser2014,Vlcek2018,Vlcek2019} The parameters of the stochastic $G_0W_0$ calculations are in Table. \ref{tab:GWsetups}.

For periodic systems, Brillouin-zone unfolding \cite{Popescu2012,Huang2014,Boykin2005,Boykin2007,Brooks2020} is performed to generate the band structure. Then many-body calculations are performed on LCB, AIB and UCB at the $k$-points accessible by the choice of the supercell to give QP energies that form QP bands. The exchange and correlation energies are extracted from our GW calculations. The bands structures are interpolated by cubic splines.

\begin{table}[H]
\centering
\begin{tabular}{c|c|c|c}
System          & Grid                                                                                  & \begin{tabular}[c]{@{}c@{}}Gridpoint Spacing\\ {(}bohr{)}\end{tabular}          & \begin{tabular}[c]{@{}c@{}}Cutoff\\ {(}hartree{)}\end{tabular} \\ \hline
1D              & 482$\times$76$\times$76                                                                             & \begin{tabular}[c]{@{}c@{}}dx=0.400664\\ dy=dz=0.4\end{tabular}                 & 26                                                             \\ \cline{2-4} 
2D              & 74$\times$144$\times$480                                                                            & \begin{tabular}[c]{@{}c@{}}dx=0.4\\ dy=0.402778\\ dz=0.401852\end{tabular}      & 26                                                             \\ \cline{2-4} 
3D              & 244$\times$72$\times$136                                                                            & \begin{tabular}[c]{@{}c@{}}dx=0.395264\\ dy=0.399167\\ dz=0.395112\end{tabular} & 26                                                             \\ \cline{2-4} 
single molecule & \begin{tabular}[c]{@{}c@{}}76$\times$76$\times$50 (F)\\ 80$\times$50$\times$56 (BT)\\ 56$\times$90$\times$96 (FBT)\end{tabular} & 0.4                                                                             & 28                                                            
\end{tabular}
\caption{Setups in the DFT calculations of all systems containing F and BT. Note: The system is periodic in x direction in 1D calculations but in y and z directions in 2D calculations.}
\label{tab:DFTsetups1}
\end{table}

\begin{table}[H]
\centering
\begin{tabular}{c|c|c|c}
System                                                                   & Grid        & \begin{tabular}[c]{@{}c@{}}Grid Spacing\\ {(}bohr{)}\end{tabular} & \begin{tabular}[c]{@{}l@{}}Cutoff\\ {(}hartree{)}\end{tabular} \\ \hline
1D fluorene                                                              & 320$\times$76$\times$76   & \begin{tabular}[c]{@{}c@{}}dx=0.397250\\ dy=dz=0.4\end{tabular}        & 26                                                             \\ \cline{2-4} 
\begin{tabular}[c]{@{}l@{}}O-substituted\\ 1D strand\end{tabular}   & 482$\times$74$\times$74   & \begin{tabular}[c]{@{}c@{}}dx=0.400664\\ dy=dz=0.4\end{tabular}        & 26                                                             \\ \cline{2-4} 
\begin{tabular}[c]{@{}l@{}}Se-substituted\\ 1D strand\end{tabular} & 482$\times$74$\times$74   & \begin{tabular}[c]{@{}c@{}}dx=0.400664\\ dy=dz=0.4\end{tabular}        & 26                                                             \\ \cline{2-4} 
1D polyacetylene                                                         & 126$\times$100$\times$100 & \begin{tabular}[c]{@{}c@{}}dx=0.298107\\ dx=dz=0.3\end{tabular}        & 26                                                             \\ \cline{2-4} 
1D polyethylene                                                          & 130$\times$100$\times$100 & \begin{tabular}[c]{@{}c@{}}dx=0.297060\\ dy=dz=0.3\end{tabular}        & 26                                                            
\end{tabular}
\caption{Setups in the DFT calculations of the other systems.}
\label{tab:DFTsetups2}
\end{table}

\begin{table}[H]
\centering
\begin{tabular}{p{0.5\textwidth}|c}
Parameter                             & Value                            \\ \hline
plane wave cut-off (hartree)                      & 26/28 (same in DFT calculations) \\
& \\
number of random vectors used for sparse stochastic compression           & 20000 \\
& \\
number of random vectors characterizing the screened Coulomb interaction (per each vector sampling the Green's function) & 15\\
& \\
number of vectors sampling the Green's function & 1000\\
& \\
maximum time for real-time propagation of the dynamical self-energy              & 50 a.u.                           
\end{tabular}
\caption{Setups in the GW calculations of all systems.}
\label{tab:GWsetups}
\end{table}

\section{Supplementary figures and tables}
\begin{figure}[H]
    \centering
    \includegraphics[width=.8\textwidth]{Fig_supercells.png}
    \caption{The supercells of periodic systems in our DFT and GW calculations. (a) A supercell of the 1D system containing 8 repeated units in the periodic direction, the polymer axis which is defined as the X direction. (b) An 8$\times$8 supercell of the 2D system that is periodic in two directions. The X direction is defined as in 1D and the Y axis denotes the polymer $\pi$-$\pi$ stacking direction. (c) A 4$\times$4$\times$4 supercell of the 3D system that is periodic in three directions. The X and Y directions are defined as in 2D. The Z direction the edge-to-edge stacking direction.} 
    \label{Fig_supercells}
\end{figure}

\begin{figure}[H]
    \centering
    \includegraphics[width=.8\textwidth]{Fig_expIPEA.png}
    \caption{Charge excitation energies, the highest occupied molecular orbitals (HOMO), and lowest unoccupied orbitals (LUMO) of the fluorene unit and the benzothiadiazole unit, respectively. Red and blue colors distinguish the wave function phase. The computed results are from the GW MB calculations, while the experimental results are available from the NIST database as indicated by the red dots on the energy axis.}  
    \label{Fig_IPEA}
\end{figure}

\begin{figure}[H]
    \centering
    \includegraphics[width=.8\textwidth]{Fig_energydiagram.png}
    \caption{Calculated QP energies and fundamental gaps of different systems. (a) Ionization potentials and electron affinities computed from GW for the FBT monomer, 1D FBT strand, and 2D FBT surface. The plotted frontier orbitals show that in the periodic systems, the valence band maximum state inherits the delocalized characteristic from the HOMO of the monomer and the conduction minimum state inherits the localized feature from the LUMO. Red and blue colors distinguish the wave function phase. (b) The fundamental gap as a function of the dimensionality of the system. Both the DFT gap and QP gap collapse as the system evolves from 0D (monomer) to 3D, while the QP gap shows a much more responsive contraction with respects to the system's topology. The red dots on the axes are available experimental results.\cite{Mai2015}} 
    \label{Fig_EnergyDiagram}
\end{figure}

\begin{figure}[H]
    \centering
    \includegraphics[width=.8\textwidth]{Fig_hybri.png}
    \caption{Formations of molecular orbitals of the FBT monomers from the donor and the acceptor units. Red and blue colors distinguish the wave function phase. The FBT HOMO retains the delocalized feature of the fluorene and the benzothiadiazole HOMO, which represents the signature of the lower conjugated band (LCB). The FBT LUMO, however, retain the acceptor LUMO only being highly localized on the acceptor unit, which is responsible for the formation of the acceptor impurity band (AIB) in the periodic system. The LUMO+1 behaves similarly to the HOMO, accounting for the formation of the upper conjugated band (UCB).} 
    \label{Fig_Hybri}
\end{figure}

\begin{figure}[H]
    \centering
    \includegraphics[width=.8\textwidth]{Fig_1Dorbitals.png}
    \caption{Selected orbitals from the bands of interest in the 1D FBT system. The periodicity of the orbital correspond to the crystal momentum of the state. Red and blue colors distinguish the wave function phase. The lower block presents 5 states from the LCB and they feature delocalized orbitals along the backbone. The middle block presents 5 states from AIB where the orbitals are highly localized on the acceptor unit regardless the change in periodicity. The upper block presents 5 states from the UCB, which are qualitatively similar to those from LCB (the lower block).} 
    \label{Fig_1Dorbitals}
\end{figure}

\begin{figure}[H]
    \centering
    \includegraphics[width=.8\textwidth]{Fig_AIB.png}
    \caption{Band structures of three D-A copolymers that have different heteratoms on the acceptor unit obtained from the Quantum-Espresso package. These three copolymers share very similar geometries (Table \ref{tab:3polymer_geometry_info}). The highlighted bands from the bottom to the top correspond to LCB, AIB, and UCB, which shows the presence of AIB regardless of the chemical modifications (highlighted in the chemical structures) on the acceptor.\cite{Chua2019}} 
    \label{Fig_AIB}
\end{figure}

\begin{table}[H]
     \centering
     \begin{tabular}{c|c|c|c|c}
 \multirow{2}{*}{system} & \multicolumn{2}{c|}{torsion angle ($^\circ$)} & \multicolumn{2}{c}{C-to-C distance (\AA)} \\ \cline{2-5} 
                         & $\phi_1$         & $\phi_2$        & $d_1$             & $d_2$                                       \\ \hline
 FBOX                    & 41.80            & 43.05           & 1.467             & 1.466                                   \\
 FBT                     & 42.37            & 43.92           & 1.470             & 1.471                                      \\
 FBSE                    & 41.68            & 42.75           & 1.471             & 1.471                                    
     \end{tabular}
     \caption{Torsion angles and C-to-C distances between the fluorene unit and three different acceptor units.}
     \label{tab:3polymer_geometry_info}
\end{table}

\begin{table}[H]
\centering
\begin{tabular}{c|c|c}
Decomposition                  & \multicolumn{2}{c}{Contribution {(}meV{)}} \\ \hline
single-electron interactions   & \multicolumn{2}{c}{-249}                   \\ \hline
classical Coulomb interactions & \multicolumn{2}{c}{1227}                   \\ \hline
exchange-correlation           & -292 (mean-field)     & -120 (non-local)    \\ \hline
total bandwidth                & 686 (DFT)             & 858 (GW)           
\end{tabular}
\caption{Individual contribution to the LCB width of 1D FBT.}
\label{tab:energydecomposition}
\end{table}

\begin{table}[H]
     \centering
     \begin{tabular}{c|ccc}
     \multirow{2}{*}{system} & \multicolumn{3}{c}{VBW GW/DFT {(}meV{)}}                                                \\ \cline{2-4} 
                         & \multicolumn{1}{c|}{x}            & \multicolumn{1}{c|}{y}            & z          \\ \hline
 \begin{tabular}[c]{@{}c@{}}1D \\ chain\end{tabular}              & \multicolumn{1}{c|}{858$\pm$(38)/686} & \multicolumn{1}{c|}{-}            & -          \\
 \begin{tabular}[c]{@{}c@{}}2D \\ surface\end{tabular}            & \multicolumn{1}{c|}{363$\pm$(26)/393} & \multicolumn{1}{c|}{626$\pm$(24)/625} & -          \\
 \begin{tabular}[c]{@{}c@{}}3D \\ solid\end{tabular}              & \multicolumn{1}{c|}{588$\pm$(33)/533} & \multicolumn{1}{c|}{711$\pm$(32)/661} & 43$\pm$(32)/23
     \end{tabular}
     \caption{Valence bandwidths (VBW) of different FBT systems in each periodic direction by DFT and GW.}
     \label{tab:bandwidths1}
 \end{table}

\begin{table}[H]
     \centering
     \begin{tabular}{c|ccc}
     \multirow{2}{*}{system} & \multicolumn{3}{c}{CBW GW/DFT {(}meV{)}}                                                \\ \cline{2-4} 
                         & \multicolumn{1}{c|}{x}            & \multicolumn{1}{c|}{y}            & z          \\ \hline
 \begin{tabular}[c]{@{}c@{}}1D \\ chain\end{tabular}              & \multicolumn{1}{c|}{265$\pm$(46)/182} & \multicolumn{1}{c|}{-}            & -          \\
 \begin{tabular}[c]{@{}c@{}}2D \\ surface\end{tabular}            & \multicolumn{1}{c|}{142$\pm$(29)/111} & \multicolumn{1}{c|}{115$\pm$(32)/55} & -          \\
 \begin{tabular}[c]{@{}c@{}}3D \\ solid\end{tabular}              & \multicolumn{1}{c|}{164$\pm$(34)/143} & \multicolumn{1}{c|}{123$\pm$(35)/132} & 302$\pm$(33)/256
     \end{tabular}
     \caption{Conduction bandwidths (CBW) of different FBT systems in each periodic direction by DFT and GW.}
     \label{tab:bandwidths2}
 \end{table}

\begin{figure}[H]
    \centering
    \includegraphics[width=.8\textwidth]{Fig_1DExP.png}
    \caption{Band structures with exchange and correlation energies of the 1D FBT system. Exchange and correlation energies are plotted relative to the band average. The $\Gamma$ to X portion stands for the transport in the polymer direction. (a) The QP energy and exchange energy are plotted  as a function of the crystal momentum for the highlighted lower and middle bands (colored). UCB is plotted in black. In LCB, the exchange grows more and more negative going from the $\Gamma$ point to the X point due to the increase in orbital overlaps, which indicates the exchange interactions broaden the valence bandwidth in the conjugated direction. In AIB, however, the exchange energy is almost insensitive to the change in states due to the fact that all the orbitals are highly localized. (b) The QP energy and correlation energy are plotted  as a function of the crystal momentum for the highlighted lower and middle bands (colored).  UCB is plotted in black. In both LCB and AIB, the correlation energy suppresses the bandwidth due to the fact that the higher the QP energy, the more negative the correlation energy (Figure ~\ref{Fig_CorrQPE}).} 
    \label{Fig_1DExP}
\end{figure}

 \begin{table}[H]
     \centering
     \begin{tabular}{c|c|c|c}
 \multirow{2}{*}{system} & \multicolumn{2}{c|}{Exchange energy (eV)} & \multirow{2}{*}{Difference (meV)} \\ \cline{2-3} 
            & Center state         & Boundary state                                                       \\ \hline
 1D PAE     & -14.59           & -16.65           & 2069 \\
 1D PEE     & -19.77           & -19.41           & -363                                 
     \end{tabular}
     \caption{The exchange energies of the states at the Brillouin center and boundary of the highest valence bands for 1D polyacetylene and polyethylene systems.}
     \label{tab:exchange-driven_band_broadening}
\end{table}

\begin{figure}[H]
    \centering
    \includegraphics[width=.8\textwidth]{Fig_CorrQPE.png}
    \caption{Correlation energy plotted as a function of the QP energy of the $\Gamma$ state (black) and the X state (blue) for systems: (a) 1D FBT strand, (b) 1D trans-polyacetylene, and (c) 1D polyethylene. The intersection between the curve and the straight line of the same color represents both the QP energy (x-coordinate) and the correlation energy (y-coordinate). All systems show the same rule that the correlation energy increases as the QP energy decreases.}
    \label{Fig_CorrQPE}
\end{figure}

\begin{figure}[H]
    \centering
    \includegraphics[width=.8\textwidth]{Fig_torsion.png}
    \caption{Molecular geometries of FBT single strands optimized in periodic systems with different dimensionalities. (a) The donor subunit and the acceptor subunit retain a rigid planar structure with unaltered bond lengths and bond angles in three systems. The main geometrical difference among the single strands is the torsion angles, $\phi_1$ and $\phi_2$, between the donor and the acceptor, which are slightly different from each other (Table \ref{tab:FBT_geometry_info}). (b) The average torsion angle ($\phi_1$$+$$\phi_2$)$/$2 of the single strands from the optimized 1D, 3D, and 2D system in the order of the magnitude being 43$^\circ$, 49$^\circ$, and 56$^\circ$.} 
    \label{Fig_torsion}
\end{figure}

\begin{table}[H]
\begin{tabular}{c|ccc}
\multirow{2}{*}{system} & \multicolumn{3}{c}{torsion angle ($^\circ$)}                                 \\ \cline{2-4} 
                        & \multicolumn{1}{c|}{$\phi_1$}  & \multicolumn{1}{c|}{$\phi_2$}  & average \\ \hline
strand of 1D chain      & \multicolumn{1}{c|}{42.37} & \multicolumn{1}{c|}{43.92} & 43      \\
strand from 3D solid    & \multicolumn{1}{c|}{49.79} & \multicolumn{1}{c|}{49.36} & 49      \\
strand from 2D surface  & \multicolumn{1}{c|}{55.67} & \multicolumn{1}{c|}{56.16} & 56     
\end{tabular}
\caption{Torsion angles between the fluorene unit and the benzothiadiazole unit in three FBT systems.}
\label{tab:FBT_geometry_info}
\end{table}

\begin{figure}[H]
    \centering
    \includegraphics[width=.8\textwidth]{Fig_2DExP.png}
    \caption{Band structures with exchange and correlation energies of the 2D FBT system. Exchange and correlation energies are plotted relative to the band average. The $\Gamma$ to Y portion stands for the transport in the polymer $\pi$-$\pi$ stacking direction while the $\Gamma$ to X portion stands for the transport along the polymer backbone. (a) The QP energy and exchange energy are plotted  as a function of the crystal momentum for the highlighted lower and middle bands (colored). UCB is plotted in black. In the polymer direction, the exchange causes the same effects as found in the 1D system, while in the $\pi$-$\pi$ stacking direction, the exchange behave oppositely. The exchange suppresses the both AIB and LCB widths. (b) The QP energy and correlation energy are plotted  as a function of the crystal momentum for the highlighted lower and middle bands (colored).  UCB is plotted in black.In both AIB and LCB and both directions, the correlation suppresses the widths due to the fact that the correlation energy increases as the QP energy decreases (Figure \ref{Fig_CorrQPE}).} 
    \label{Fig_2DExP}
\end{figure}

 \begin{table}[H]
     \centering
     \begin{tabular}{c|c|c|c|c}
 \multirow{2}{*}{supercell size} & \multicolumn{2}{c|}{HOMO {(}eV{)}} & \multicolumn{2}{c}{Transport gap {(}eV{)}} \\ \cline{2-5} 
                                 & DFT          & GW                  & DFT              & GW                      \\ \hline
 2$\times$2$\times$1                           & -4.51        & -5.12$\pm$(0.03)        & 1.52             & 3.02$\pm$(0.05)             \\
 4$\times$4$\times$1                           & -4.53        & -5.52$\pm$(0.03)        & 1.52             & 3.34$\pm$(0.04)             \\
 6$\times$6$\times$1                           & -4.54        & -5.45$\pm$(0.02)        & 1.45             & 3.31$\pm$(0.04)             \\interactions
 8$\times$8$\times$1                           & -4.54        & -5.48$\pm$(0.02)        & 1.52             & 3.33$\pm$(0.03)            
 \end{tabular}
     \caption{Convergence of HOMO energy levels and transport gaps to the  supercell size in 2D calculations.}
     \label{tab:convergence_to_supercell}
 \end{table}

\bibliography{MB_vdW_OC}